\documentclass[twocolumn]{svjour3}          % twocolumn
\journalname{}%Granular Matter}
\usepackage{amsmath} % for bold greek fonts

\usepackage{graphicx}        % standard LaTeX graphics tool
\usepackage{epstopdf}

\begin{document}

\title{The interactions between comminution, segregation and remixing in granular flows}

\author{Benjy Marks \and Itai Einav}
\institute{B. Marks \at
           Particles and Grains Laboratory, The University of Sydney \\
           \email{benjy.marks@sydney.edu.au} \and
           I. Einav \at
           Particles and Grains Laboratory, The University of Sydney}
\date\today

\maketitle

\begin{abstract}
Granular segregation is an important mechanism for industrial processes aiming at mixing grains. Additionally, it plays a pivotal role in determining the kinematics of geophysical flows. Because of segregation, the grainsize distribution varies in space and time. Additional complications arise from the presence of comminution, where new particles are created, enhancing segregation. This has a feedback on the comminution process, as particles change their local neighbourhood. Simultaneously, particles are generally undergoing remixing, further complicating the segregation and comminution processes. The interaction between these mechanisms is explored using a cellular automaton with three rules: one for each of segregation, comminution and mixing. The interplay between these rules creates complex patterns, as seen in segregating systems, and depth dependent grading curves, which have been observed in avalanche runout. At every depth, log-normal grading curves are produced at steady state, as measured experimentally in avalanche and debris flow deposits.

\keywords{Segregation \and Comminution \and Mixing \and Cellular automata}
%\PACS{PACS code1 \and PACS code2 \and more}
%\subclass{MSC code1 \and MSC code2 \and more}
\end{abstract} % 250 word limit

%\begin{keyword}
%Segregation \sep Comminution \sep Cellular automata \sep Debris flow
%\end{keyword}

%\end{frontmatter}

\section{Introduction}

\begin{figure}
    \centering
    \includegraphics[width=\columnwidth]{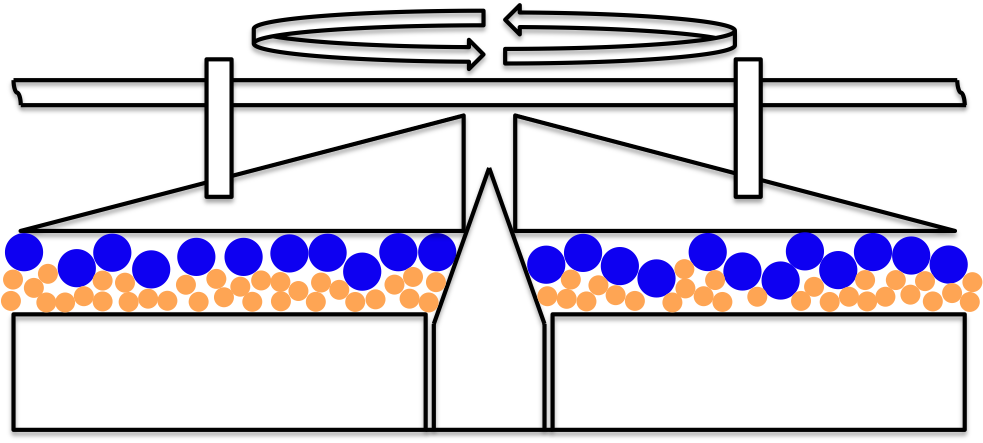}
    
    \vspace{5pt}
    \includegraphics[width=\columnwidth]{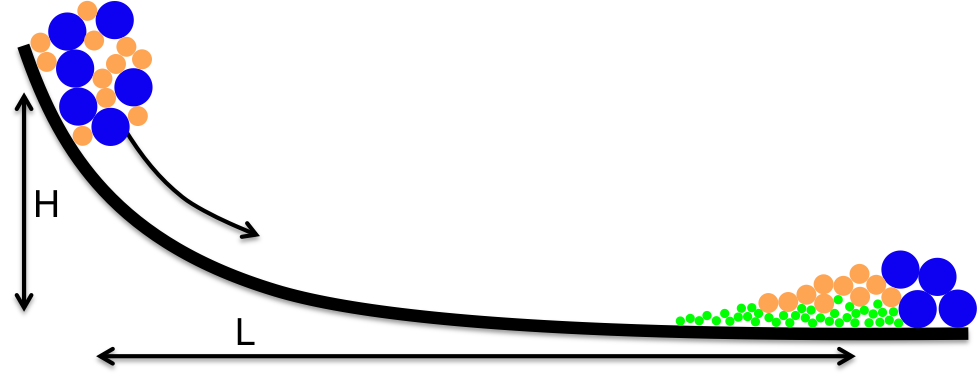}
    \caption{\emph{Top}: A mill stone --- the grinding of grain using a mill stone is one of the oldest industrial problems in human history, yet still mathematically unsolved. Particles of grain are crushed to a fine powder by very large deformation shearing at high normal stress. The fine powder segregates out of the shear zone into cavities built into the mill stone, and then under the action of centripetal forces migrates out to a collection bin. \emph{Bottom}: A long run-out landslide, where the ratio of $L/H$ can be up to 10. $L$ and $H$ are the change in horizontal position and height respectively of the centre of mass of an avalanche during run-out. Large values of $L/H$ are possible indicators of lubrication by a layer of very small particles at the base of the flow, which have been created as a result of comminution and percolated downwards through segregation. \label{fig:mill_stone}}
\end{figure}

Flowing granular material exhibits complex behaviour, including many phenomena that cannot be described in the context of a conventional fluid. For example size segregation, comminution and agglomeration all have no analogue in traditional fluids. To describe these phenomena the grainsize distribution has o be involved as a dynamic property.

The dynamics of granular material are important in many natural processes, such as debris flows, landslides, rockfalls and shear banding. Industrial processing also requires many granular flows, such as rotating or tumbling mills, chute flows and hopper filling or discharge. In these types of flow processes, breakage of particles can be important. As yet, there are no temporo-spatial continuum models for measuring changes in the grainsize distribution for such open systems where particles can advect in space.

This is especially important in two poorly understood systems, shown in Figure \ref{fig:mill_stone}. The first is ancient grain milling, where combined normal and shear stresses crush wheat grains, dynamically sieving the resultant mix. The second is long run out avalanches, where it is unclear why these incredibly destructive natural phenomena can travel enormous distances, up to 10 times their vertical fall \cite{Legros2002301,imre2010fractal}. 

With regards to natural flows, it has been understood for some time that there is a need to model spatial and temporal variability in the grainsize distribution of flowing material to be able to implement appropriate rheological models \cite{iverson2003debris}.

Here we tackle this problem using a cellular automaton \cite{Wolfram} with three distinct rules of operation: segregation, remixing and comminution. As has been shown previously \cite{sizeism} for bidisperse systems, we can describe segregation in terms of the swapping of cells in a cellular automaton. Comminution rules have also been developed \cite{steacy1991automaton,mcdowell1996fractal}, but these are limited to closed systems. Here we will present rules which can apply in open systems to arbitrarily polydisperse materials.

As in \cite{grainsize} we denote the grainsize, $s$, as an internal coordinate of the system such that every point in space has a grainsize distribution. We then describe this continuous grainsize distribution $\phi(s)$ of the system in terms of the solid fraction $\Phi(s)$ of particles between grainsizes $s_a$ and $s_b$ as

\begin{equation}
\Phi[s_a<s<s_b] = \int_{s_a}^{s_b}\phi(s')~ds'.
\end{equation}

Conservation of mass at a point in space ${\bf r} = \{x,y,z\}$ can then be expressed as \cite{grainsize,redner1990fragmentation,ramkrishna2000population}

\begin{equation}\label{eq:mass}
\frac{\partial\phi}{\partial t} + {\bf \nabla}\cdot(\phi{\bf u}) = h^+ - h^-,
\end{equation}

\noindent where ${\bf u}({\bf r},s,t)=\{u,v,w\}$ is the material velocity, \linebreak$h^+({\bf r},s,t)$ is the birth rate, describing the creation of new particles of grainsize $s$ at time $t$, and $h^-({\bf r},s,t)$ is the death rate, at which particles of grainsize $s$ are destroyed.

%\begin{figure}
%    \centering
%    \includegraphics[width=\columnwidth]{images/breakage_rule_pcolor}
%    \caption{A small example using the breakage rule on 10 cells over time.\label{fig:break_small}}
%\end{figure}

The second term in Equation \ref{eq:mass} describes the advection of mass, such as characterises open systems, where material can move spatially. The right hand side of the same Equation represents mechanisms traditionally treated as closed systems, such as agglomeration, crushing and abrasion. Each of these systems --- open and closed --- has been the subject of much study, but the coupling of such processes using a continuum description has yet to be achieved.

\section{Cellular automata and continua}

To model these systems we use a cellular automaton in two spatial dimensions. It is a series of cells on a 2D regular cartesian lattice  of size $N_x$ by $N_z$, with directions $x$ and $z$, and cell spacing $\Delta x$ and $\Delta z$ in the respective directions. The $z$ direction is perpendicular to the shear direction, such that for inclined plane flow it points normal to the slope. The $x$ direction represents a micro scale internal coordinate.

We allow the grainsize distribution to be a function of this internal spatial coordinate such that $s=s(x)$. By discretising the grainsize distribution $\phi(s)$ into $N_x$ monodisperse components of equal volume, they can be arranged in the $x$ direction such that summing over this coordinate would recover the full grainsize distribution. We consider the local neighbourhood of a particular particle as those that are adjacent in the $x$ direction. The $x$ direction now contains more information than the grainsize distribution alone, as the local orientation of particles is preserved, below the resolution of the analogous continuum scale.

We number the cells from the bottom-left corner of the grid so that position on the grid can be expressed using the pair $\{i,j\}$, where $i$ and $j$ indicate the number of cells across in the respective $z$ and $x$ directions. In all cases the system is considered to be periodic in the $j$ direction.

Each cell contains a single number, $s_{i,j}$, which dictates the grainsize of the particles in the representative volume element defined by the cell $\{i,j\}$.

We can define a discretised grainsize distribution $\phi_i$ at any height $i$ as a histogram of the number of cells within a discrete grainsize fraction with centre $s_a$ and width $\Delta s$ in all $N_x$ neighbours taken in the $j$ direction:

\begin{equation}
\phi_{i}(s_a) = \frac{1}{N_x\Delta s}\sum_{k=1}^{N_x}
\begin{cases}
    1 &\text{if } s_a-\frac{\Delta s}{2}<s_{i,k}\le s_a+\frac{\Delta s}{2},\\
    0 &\text{otherwise}.
\end{cases}
\end{equation}

We also define the local average grainsize over the nearest neighbours in the $j$-direction as

\begin{equation}\label{eq:s_bar}
\overline s_{i,j} = (s_{i,j-1} + s_{i,j+1})/2,
\end{equation}

\noindent and similarly for the $i$-direction. For the cellular automata rules defined below, we need to define two new operators: $s_a \Leftrightarrow s_b$ represents the swapping of values $s_a$ and $s_b$ between their respective cells, and $s_a \Rightarrow s_b$ represents a change in grainsize in the cell containing $s_a$ to the new value of $s_b$.

\begin{figure}
    \centering
    \includegraphics[width=\columnwidth]{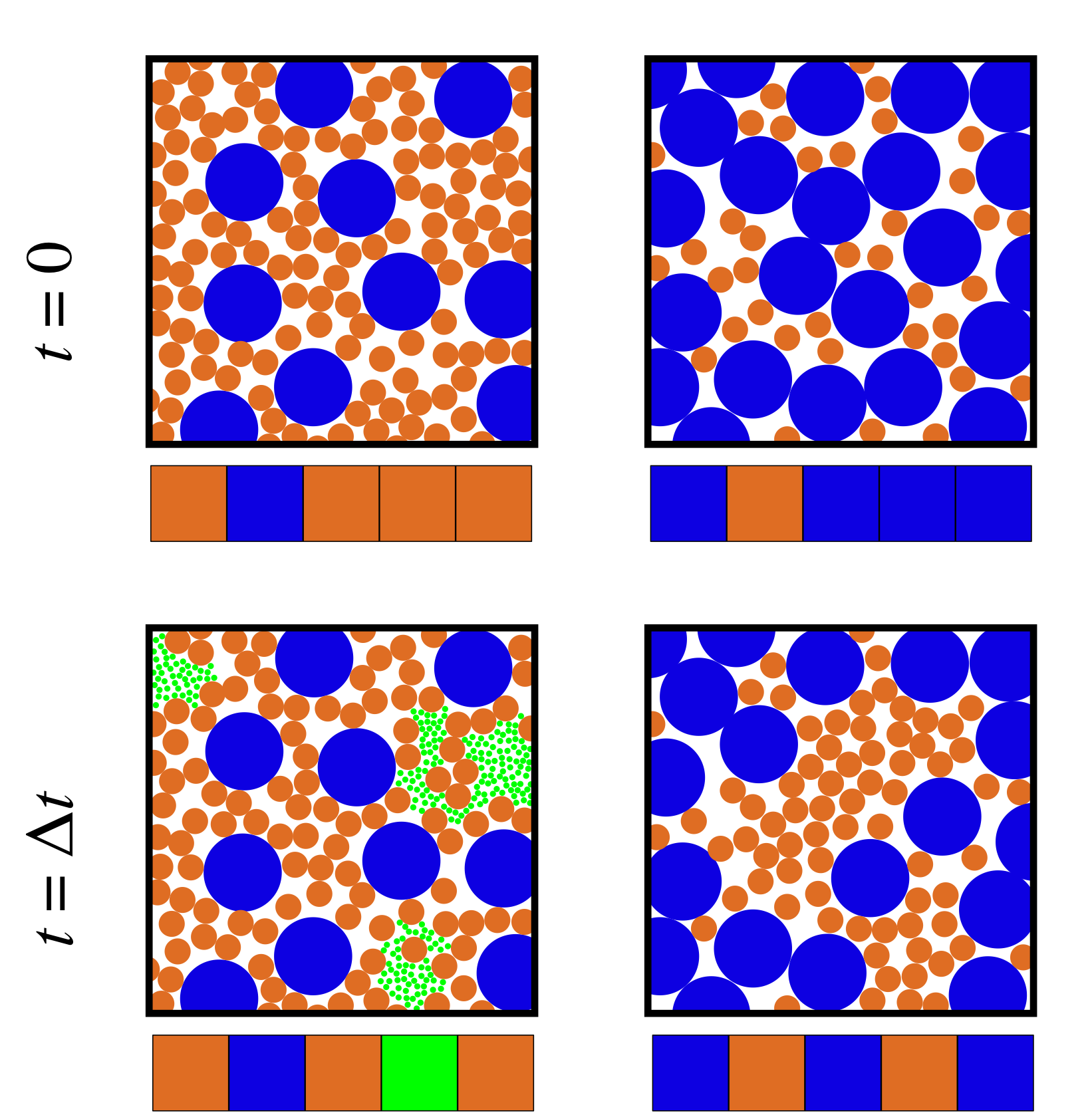}
    \caption{The cushioning effect and nearest neighbour rule. \emph{Left:} Large particles are cushioned such that they will not break because of an abundance of small particles. \emph{Right:} Small particles do not carry a significant amount of load as they are free to move in interstitial pore spaces. \emph{Top:} Initial conditions. \emph{Bottom:} Result after one iteration of the cellular automata rule, where on the left the small particles in a cell (orange) become even smaller (green), and on the right the large particles in a cell (blue) become smaller.\label{fig:cushioning}}
\end{figure}

In the limit of infinitesimal cell size, the cellular automaton may be regarded as a first order partial differential solver \cite{toffoli1984cellular}. In fact, there are many ways to recover continuous results from such a cellular automaton \cite{smoothlife}.

\section{Closed systems}\label{sec:breakage}

We begin by considering closed systems, which are those in which material does not advect in space such that ${\bf u} = {\bf 0}$. There are many processes that have previously been represented as closed systems, such as comminution, agglomeration and abrasion. We here look only at comminution, where particles are crushed to form fragments of smaller sizes.

Following the formulation in \cite{ramkrishna2000population}, for the case where particles are created only by the fragmentation of larger particles, we can express the death rate as

\begin{equation}
h^-(s,t) = b(s)\phi(s,t),
\end{equation}

\noindent where $b(s)$ is some specific breakage rate which governs the frequency at which particles of grainsize $s$ break into smaller fragments. The birth rate is then the sum of breakages into size $s$ over all particles larger than $s$, which can be expressed as 

\begin{equation}
h^+(s,t) = \int_s^\infty b(s')P(s|s')\phi(s',t)~ds',
\end{equation}

\noindent where $P(s|s')$ is a probability density function which dictates the probability of creating grainsize $s$ from crushing a particle of grainsize $s'$. We can then express conservation of mass as

\begin{equation}\label{eq:cons_closed}
\frac{\partial\phi(s,t)}{\partial t} = \int_s^\infty b(s')P(s|s')\phi(s',t)~ds' -b(s)\phi(s,t).
\end{equation}

In a discrete sense, such as that defined in the cellular automaton, we can rewrite this equation as the conservation of a grainsize fraction with centre $s_a$ and width $\Delta s$, over a time step $\Delta t$ as

\begin{align}\label{eq:cons_closed_discrete}
%\frac{\Delta\phi_i(s_a)}{\Delta t} = \sum_{j=1}^{N_x}\Bigg(\sum_{k=a+\Delta s}^{N_s} &\Big(b_i(s_k)P(s_a|s_k)\phi_i(s_k)\Delta s\Big) - \nonumber\\
%&b_i(s_a)\phi_i(s_a)\Bigg),
\frac{\Delta\phi_i(s_a)}{\Delta t} = \sum_{k=a+\Delta s}^{N_s} \Big(&b_i(s_k)P(s_a|s_k)\phi_i(s_k)\Delta s\Big) - \nonumber\\ &b_i(s_a)\phi_i(s_a),
\end{align}

\noindent where $N_s$ is the total number of evenly spaced bins of size $\Delta s$. These equations have been considered many times before \cite{randolph1977effect,peterson1985comparison,mcgrady1987shattering,williams1990exact}, and solutions have been proposed for many mechanisms of comminution, such as grinding, cleavage and abrasion. However, breakage mechanisms are normally assumed with a priori knowledge of power law distributions \cite{redner1990fragmentation,randolph1977effect}. In fact, in most models, either the breakage rate $b$, the fragment probability distribution $P$, or both, are generally assumed to be power law in nature from the outset \cite{randolph1977effect}.

Another method to model the problem has been proposed in various forms, and uses simple geometric analogies in a cellular automaton \cite{steacy1991automaton,mcdowell1996fractal} where power law patterns are found, not imposed, by assuming that particles with neighbours of the same size are likely candidates to crush.

We can unify these two approaches, of macroscopic grainsize distribution changes, and microscopic nearest neighbourhood behaviour, by including the grainsize distribution in a cellular automaton.

\subsection{The crushing mechanism}

When a large particle is surrounded by small particles, it is cushioned from fracture by having many points of contact with neighbours, leading to a fairly isotropic loading state, as shown on the left of Figure \ref{fig:cushioning}. Also, when a small particle is surrounded by large particles, it does not carry significant load, as it is either able to fit in the pore spaces if sufficiently small, or is highly mobile, as on the right of Figure \ref{fig:cushioning}. Because of this, we consider fracture of particles only when they are surrounded by particles of a similar size. We then have a condition for fracture such that for a particle of size $s_{i,j}$ and a local neighbourhood of particles of average grainsize $\overline s_{i,j}$,

\begin{equation}\label{eq:fracture_condition}
s_{i,j} \Rightarrow (0,1)\times s_{i,j}, \hspace{20pt}\text{if } |s_{i,j} - \overline s_{i,j}| \le \beta s_{i,j}.
\end{equation}

\begin{figure}
    \centering
    \includegraphics[width=\columnwidth]{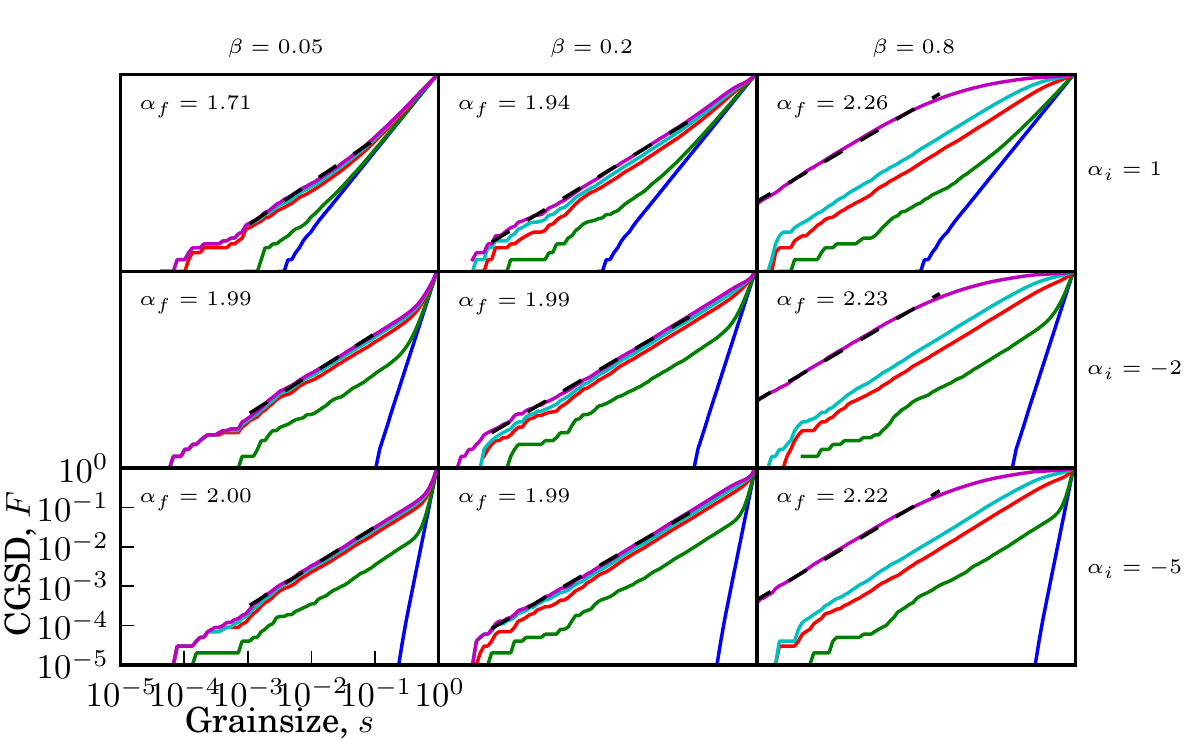}
    \caption{Evolving cumulative grainsize distributions due to comminution. All simulations have $N_z=1$, $N_x = 10^6$ and $k_b = 1$. Three initial conditions are considered, with $\alpha_i =$ --5, --2 and 1. Each initial condition is crushed at three values of $\beta = $ 0.05, 0.2 and 0.8. Within each sub-plot, each line represents a different time $t=0$, 0.05, 1, 2 and 10, from bottom right to top left.\label{fig:break}}
\end{figure}

This rule states that all particles in the representative volume defined at $s_{i,j}$ reduce in size by a randomly chosen factor between 0 and 1 if it is within $\beta s_{i,j}$ of the local mean grainsize. The non-dimensional factor $\beta$ determines how similar particles must be to their neighbours before crushing can occur. We have also included a factor of $s_j$ on the right hand side of the inequality to make small particles harder to crush, recognising that smaller particles have a higher crushing stress than larger ones \cite{Weibull}. This crushing event has some frequency which is proportional to the shear strain rate, allowing us to define the breakage rate as

\begin{equation}
b_{i,j} = k_b|\dot\gamma_i|\mathcal{H}\left(\beta s_{i,j} - |s_{i,j} - \overline s_{i,j}|\right),
\end{equation}

\noindent where $k_b$ is a non-dimensional fitting parameter and $\mathcal{H}$ is the Heaviside function. In (\ref{eq:fracture_condition}) we have also defined the fragment size distribution as 

\begin{equation}
P(s_a|s_{i,j}) = \Delta s/s_{i,j},
\end{equation}

\noindent where $\Delta s$ is the size of the grainsize bin in the $s$ direction containing fragment size $s_a$. The simulation occurs over $N_x$ cells, spaced $\Delta x$ apart. Initial conditions are generated by sampling $N_x$ times from $F=\Delta x$ to $F=1$ linearly along the inverse power law distributions defined by

\begin{equation}\label{eq:initial}
s=F^{1/(3-\alpha_i)}.
\end{equation}

\noindent where $F(s)=\int_{s_m}^s\phi(s')~ds'$ is the cumulative grainsize distribution function and $3-\alpha_i$ is the power law gradient. By time marching with a sufficiently small time step, such that $b_i\Delta t \le 1$, we implement the frequency of breakage in a probabilistic manner.

\begin{figure}
    \centering
    \includegraphics[width=\columnwidth]{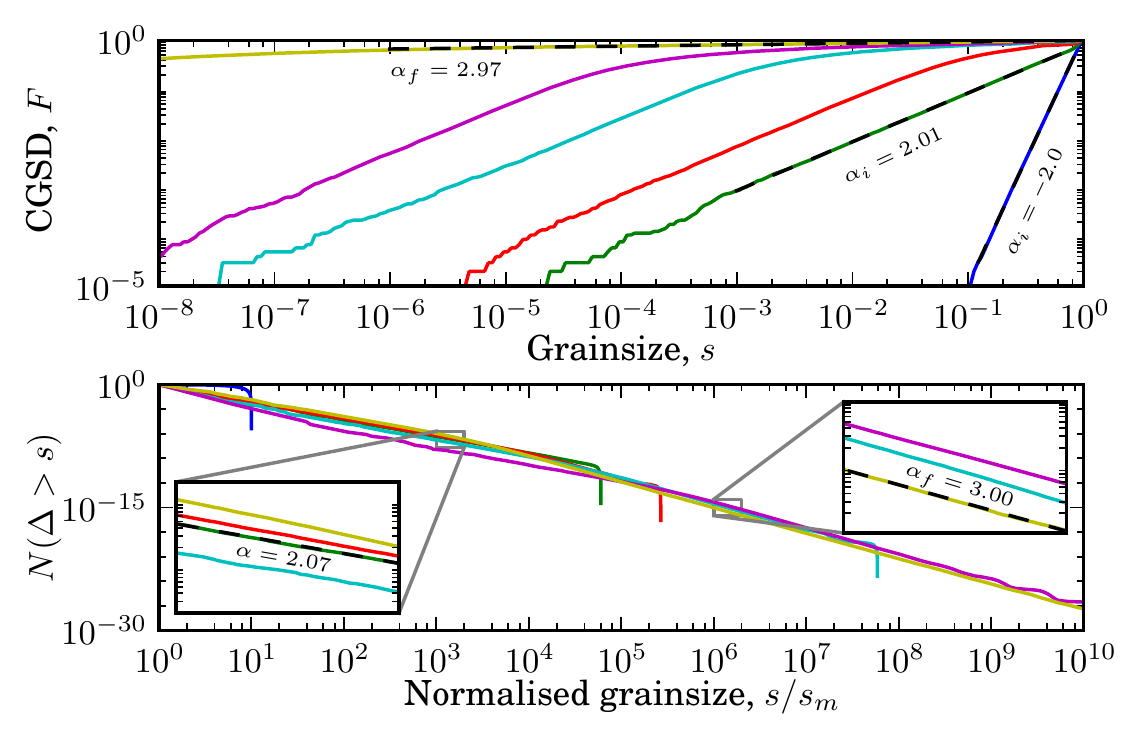}
    \caption{Cycles of breakage towards an attractor. \emph{Top:} Cumulative grainsize distributions at different numbers of cycles of loading. Each line represents the grading after a certain number of crushing-remixing cycles. From bottom right to top left, they are 0, 1, 10, 50, 100 and 500 cycles. The system approaches $\alpha=3$, or a horizontal line in this plot. \emph{Bottom:} The same data as in the top plot, now shown as number of particles $\Delta$ greater than a certain size $s$. Insets show zoomed areas of original plot.\label{fig:cycles}}
\end{figure}

Figure \ref{fig:break} shows three different initial conditions, and their progression to a steady state grainsize distribution. Each initial condition is considered for three values of $\beta$. For each case, a new distribution is reached after a single timstep of $\Delta t=0.05$. As time progresses, $\beta$ controls most of the behaviour of the process. For small values of $\beta$, a steady state is reached where the grainsize distribution does not change appreciably after time $t=5$, resulting in a power law dimension of $\alpha = 1.99\pm0.01$. For large $\beta$, all of the largest particles are continually crushed, eventually resulting in a system where $\alpha$ approaches 3.

The cellular automata predicts the same final grainsize distribution largely independent of the initial grading, with some minor effects due to the initial concentration of very large particles. Such a power law grading of the grainsize distribution has been measured in other cellular automata \cite{steacy1991automaton,mcdowell1996fractal}, discrete element simulations \cite{OdedForceAttractor} and experiment \cite{imre2010fractal}.

Generally fractal dimensions are measured in fault gauges, confined comminution tests and rock avalanches in the range of $\alpha=2$ -- 3 \cite{benzion2003,turcotte1986fractals,crosta2007fragmentation}. The limiting value of $\alpha=2$ for most cases of our model can be explained using the cellular automaton developed in \cite{turcotte1986fractals}, where every particle in the systems has the same probability of crushing, given some additional geometric constraints. If this probability is exactly 0.5, the system develops a fractal dimension of $\alpha=2$. If the probability is 1, the system reaches $\alpha=3$, which represents a system with large strain, where nearest neighbours change over the crushing period \cite{sammis2007mechanical}. This distribution in fact corresponds to a random appollonian packing, as in \cite{delaney2008relation}.

\begin{figure}
    \centering
    \includegraphics[width=\columnwidth]{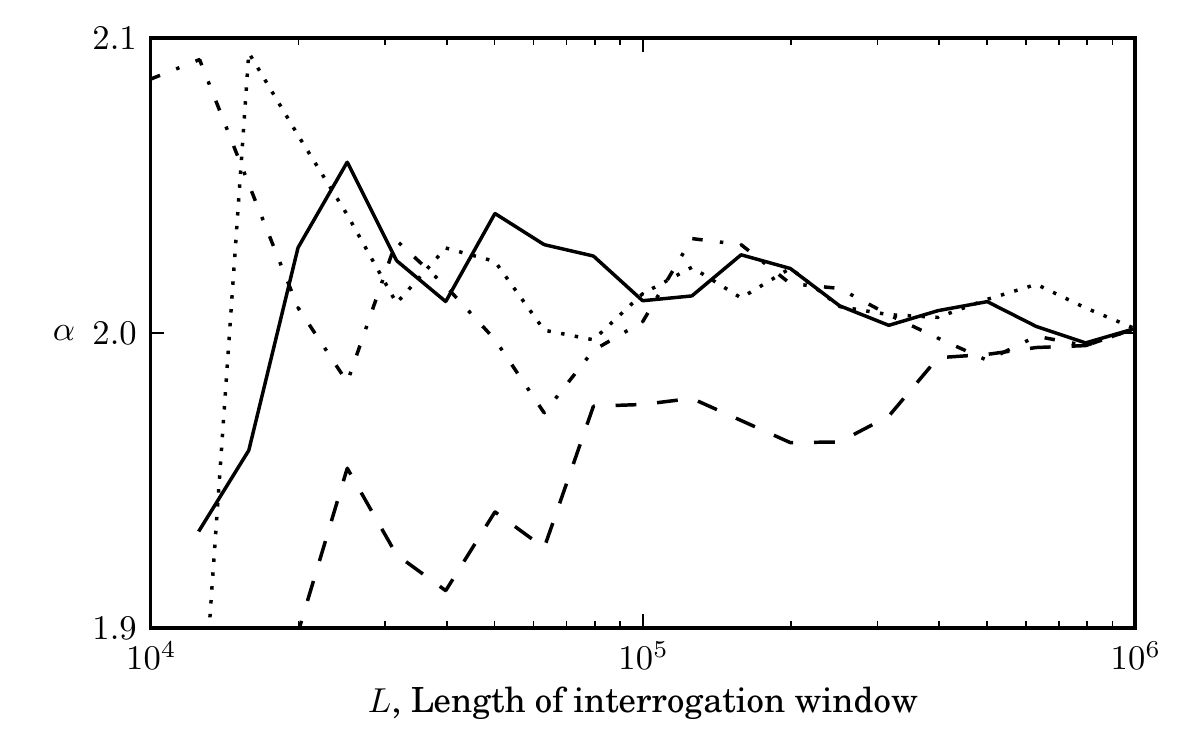}
    \caption{Fractal dimension $\alpha$ against interrogation window length $L$. $\alpha$ value is measured as a best fit of the cumulative grainsize distribution from $s=10^{-4}$ to $10^{-1}$, restricted to the cells between $j=a\pm L/2$. Each line represents a different value of $a=0,N_x/4,N_x/2$ and $3N_x/4$. For this simulation $N_x=10^6$, $\beta=0.2$, $k_b=1$ and $\alpha_i=-2$. \label{fig:box_count}}
\end{figure}

In Figure \ref{fig:box_count} $\alpha$ is measured many times in a single simulation. A number of cells, $L$, centred at $a$, are chosen, and a best fit of $\alpha$ is measured for the cumulative grainsize distribution. With increasing zoom into the system, the measured power law does not change slope significantly. This is an indication of a fractal distribution \cite{steacy1991automaton}.

The idea of such final power law grainsize distributions has been applied in \cite{BreakageMechanics} to develop a thermodynamically consistent theory for comminution processes in granular materials in closed systems. Extension of this or other theories to open systems where particles can advect has not yet been achieved.

\subsection{Comparison with continua}

To compare the cellular automata rules with a continuum, we consider the evolution of a large number of cells simultaneously, and find averaged properties that represent the continuum scale. We express the breakage rate, $b_i$, of a single grainsize fraction $s_a$ covering sizes over a range of $\Delta s$ as

\begin{equation}\label{eq:b_average}
b_i = \frac{k_b|\dot\gamma_i|}{N_x}\sum_{j=1}^{N_x}\mathcal{H}\left(\beta s_{i,j} - |s_{i,j} - \overline s_{i,j}|\right).
\end{equation}

%Together with (\ref{eq:cons_closed}), the evolution of the grainsize distribution $\phi_i$ over a single time step can then be expressed in terms of these rates as

%\begin{align}\label{eq:crush_continuum}
%&\frac{\Delta\phi_i(s_a)}{\Delta t} = k_b|\dot\gamma_i|\Bigg(- \phi_i(s_a)\sum_j \mathcal{H}(\beta s_k - |\overline s_{i,j} - s_k|) +\nonumber\\
%&\sum_{k=a+\Delta s}^{N_s}\Big(\phi_i(s_a)\frac{\Delta s}{s_k}\sum_j \mathcal{H}(\beta s_k - |\overline s_{i,j} - s_k|)\Big)\Bigg)
%\end{align}

To solve this system globally, we need to sum over $j$, which represents local information about the nearest neighbours. If we were to represent this in a continuum sense, this local information is smaller than the continuum scale, and so a new length scale, $\zeta$, must be introduced. In the cellular automaton, the length scale which controls this behaviour is that of the nearest neighbour zone (here set arbitrarily to $\Delta x$). By increasing the size of the neighbourhood over which we find the local average grainsize, the system would converge towards a different state, where physical proximity of neighbours is not considered.

\begin{figure}
    \centering
    \includegraphics[width=0.85\columnwidth]{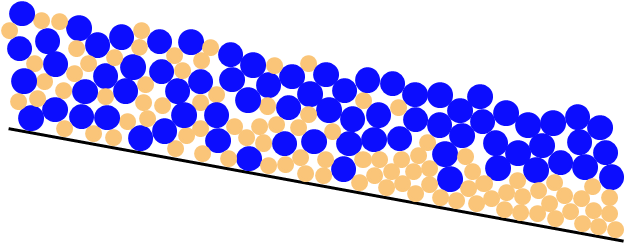}
    \includegraphics[width=0.85\columnwidth]{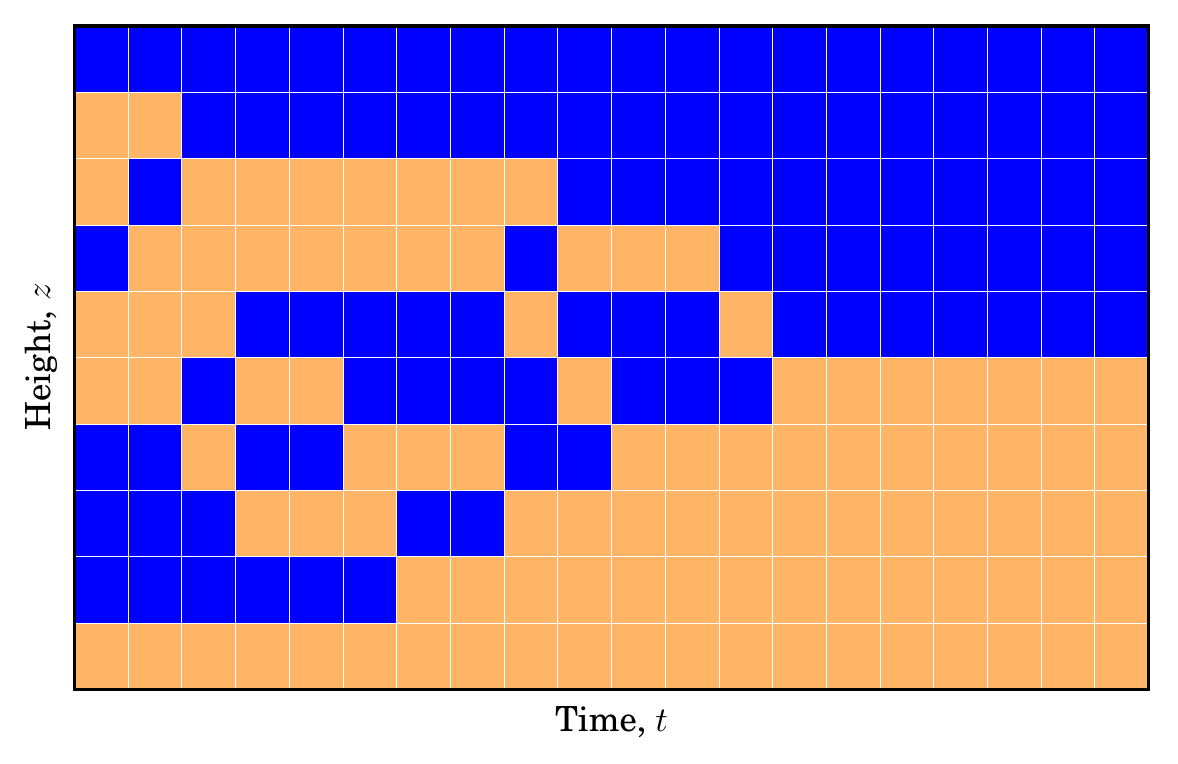}
    \caption{Bidisperse cellular automaton. \emph{Top:} Schematic representing bidisperse segregation in 2D flow down an inclined plane. \emph{Bottom:} The complimentary 1D cellular automaton, where large particles and small particles swap over time. \label{fig:bi_schematic}}
\end{figure}

How to introduce such a length scale in a continuum theory is an open question. In the cellular automaton, we can consider changing this length scale by doing cycles of crushing, as will be shown below.

\begin{figure}
    \centering
	\includegraphics[width=\columnwidth]{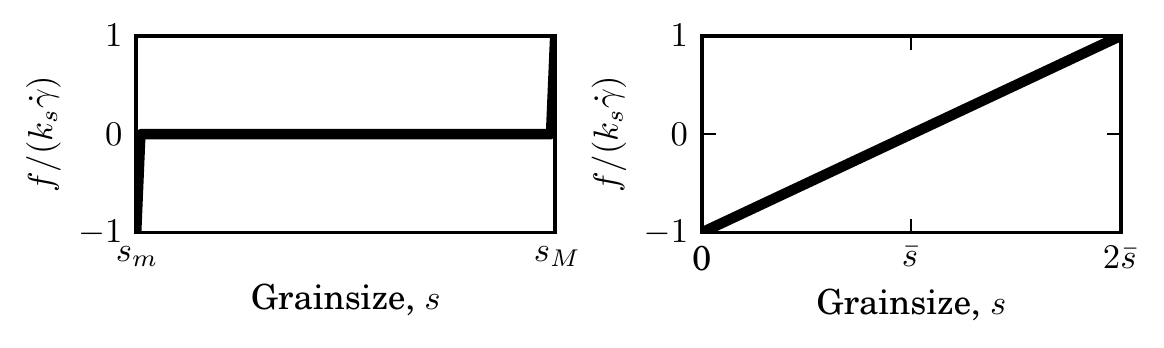}
    \caption{The segregation mechanism as a function of gransize $s$. \emph{Left:} Bidisperse rule used previously in \cite{sizeism} \emph{Right:} Polydisperse rule used here, where segregation frequency is a function of distance to the local mean size. \label{fig:f_s}}
\end{figure}

\subsection{Cycles of crushing: Towards open systems}

We can extend this simulation to a quasi-open system by considering a rearrangement of particles within the cellular automaton, but without true advection. 

After the system has reached steady state, and no further significant crushing will occur, we shuffle the system, relocating all of the nearest neighbours, and resume crushing until a new steady state is reached. We can continue this cycle until an \emph{ultimate} steady state is reached.

We begin with the simulation shown in the centre of Figure \ref{fig:break}, which has an initial grading defined by $\alpha_i=-2$, and after one full crushing iteration process has reached $\alpha=2$. As progressively more iterations occur, a region of higher slope develops at larger grainsizes, and this propagates to lower grainsizes with increasing iterations, as shown at the top of Figure \ref{fig:cycles}. This trend is towards $\alpha=3$, however this coincides with a horizontal line in this plot, and cannot be seen clearly. To visualise this more clearly, the bottom plot in Figure \ref{fig:cycles} shows the same data, but plotted as the number of particles $\Delta$ greater than a certain size $s$, such that the slope of the graph is $\alpha$. We define the number of particles per cell as being inversely proportional to the size cubed. The final state has been shuffled and crushed 500 times, tending towards an ultimate grading with this new power law gradient of $\alpha_f=3$.

This effect of cycles of crushing towards $\alpha_f=3$ has been observed experimentally \cite{lHorincz2005grading}, numerically with a crushable discrete element method \cite{oded2010confined} and predicted analytically as the maximum entropy path towards the least efficient packing of the system \cite{delaney2008relation}. It is remarkable that such a simple, one dimensional system as this can replicate the packing involved in such a complex system.

\begin{figure}
    \centering
    \includegraphics[width=\columnwidth]{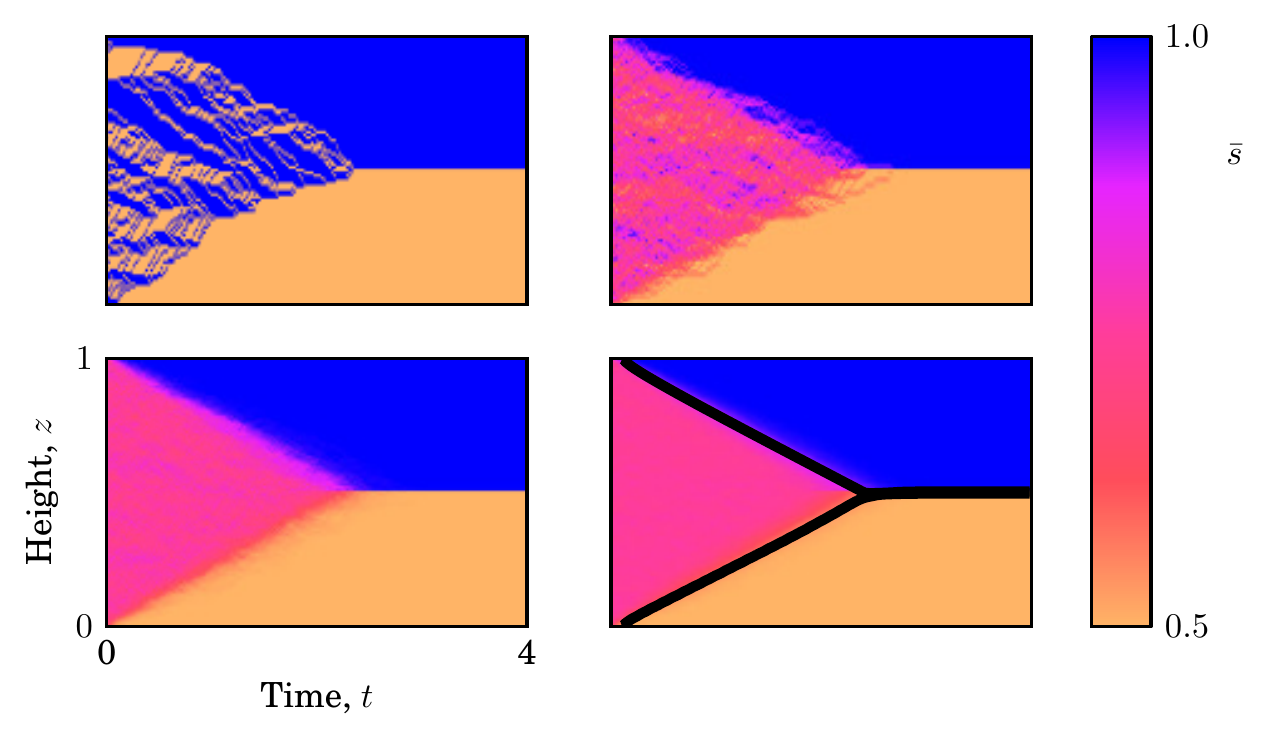}
    \caption{The time evolution the average grainsize $\bar s$ of a bidisperse simulation with varying $N_x$ subject to segregation only. All cases have $k_s=1$ and $N_z=100$. \emph{Clockwise from Top Left:}  $N_x=1$, 10, 100 and 1000. For the case of $N_x=1$ we find the local average grainsize in the $z$ direction. \emph{Bottom right:} Black lines indicate positions of concentration shocks from solution of analogous continuum equation.\label{fig:averaging}}
\end{figure}

\section{Open systems}

In order to model open systems, where particles can advect between points in space, we need rules for advection. To be a truly physical model, we need to satisfy the conservation equation for grainsize $\phi$ such that

\begin{equation}
\frac{\partial\phi}{\partial t} + {\bf \nabla}\cdot(\phi{\bf u}) = 0.
\end{equation}

We set up our rules for the cellular automaton such that this conservation law is held for some velocity ${\bf u}$. The first mechanism we consider is due to segregation. Towards this end, we begin with the simplest case of segregation and describe bimixtures, where we present a model similar to that proposed in \cite{sizeism}.

\subsection{The segregation mechanism}

As particles flow they collide, creating new void spaces which are preferentially filled by smaller particles moving, as in Figure \ref{fig:bi_schematic}. The rate of creation of void spaces is governed by the shear strain rate, $\dot\gamma$.

The simplest description of this system in terms of grainsize is shown on the left of Figure \ref{fig:f_s} for a bimixture of sizes $s_m$ and $s_M$, where the swapping frequency $f$ is defined such that small particles always move down, and large particles move up.

We can extend this model to describe polydisperse materials by using the formulation developed in \cite{grainsize}, where it was shown through energy considerations that if a particle is larger than the local average, it has some probability of moving up, and conversely if it is smaller than the average it will move down. Increasing distance in the $s$ direction from the average will increase the likelihood of swapping linearly. This is shown on the right of Figure \ref{fig:f_s}.

\begin{figure}
    \centering
    \includegraphics[width=\columnwidth]{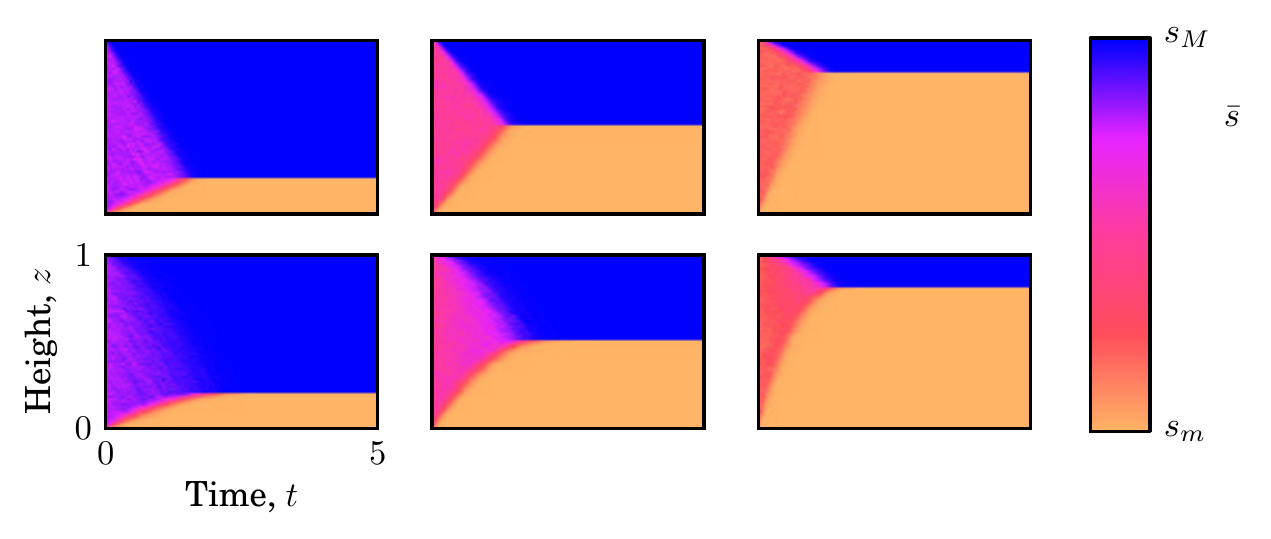}
    \caption{Time evolution for bidisperse shear flows subject to segregation. For all cases $k_s=1$. \emph{Left to right}: The system is initially filled with 20\%, 50\% and 80\% small particles respectively. \emph{Top row:} Simple shear, where $\dot\gamma=1$. \emph{Bottom row:} Inclined plane flow shear condition, where $\dot\gamma=\sqrt{1-z}$.\label{fig:bi_seg_both}}
\end{figure}

We facilitate this movement by swapping this grainsize with that either above, $s_{i+1,j}$ or below, $s_{i-1,j}$ depending on whether it is smaller or larger than the average size $\overline s_{i,j}$. We define the rate of swapping as $f =k_s|\dot\gamma_{i}|(s_{i,j}/\overline s_{i,j} - 1)$, with the sign determining the direction:

\begin{eqnarray*}
\text{if }
\begin{cases}
    f > 0,& s_{i,j} \Leftrightarrow s_{i+1,j}\\
    f < 0,& s_{i,j} \Leftrightarrow s_{i-1,j}.
\end{cases}
\end{eqnarray*}

\noindent where $k_s$ is a non-dimensional parameter controlling the rate of segregation. We iterate in two half time steps, alternately applying this rule firstly to all odd rows, and then all even rows, so that particles are inhibited from moving very large distances in a single time step.  Additionally, we only allow swapping upwards if the particle is larger than the one above it, or smaller than the one below it if moving downwards.

\subsection{Bidisperse segregation}\label{sec:bi}

To model a simple bidisperse material, we take a single column of a bimixture ($N_x=1$), of equal proportions of sizes $s_m$ and $s_M$, randomly allocated to cells, and allow it to segregate under simple shear with $\dot\gamma(z)=1$ and $k_s=1$. The result is shown in the top left of Figure \ref{fig:averaging}. We can then run the simulation with $N_x=5$ and average over the $x$-direction. The result of this is shown in the top right of Figure \ref{fig:averaging}. We can do this repeatedly, to get an increasingly resolved image of the process, as shown in the bottom row of Figure \ref{fig:averaging} for $N_x=50$ and 1000. With increasing resolution, this converges on the analytic solution presented in \cite{grainsize} and \cite{gray2005tps}.

Figure \ref{fig:bi_seg_both} shows the average grainsize at each height over time for two different shear regimes, each for 3 different initial concentrations of small particles, $s_m$. The top row depicts simple shear, as in Figure \ref{fig:averaging}, while the bottom row uses a simplified version of the shear strain rate profile predicted in \cite{grainsize} for the case of inclined plane flow: $\dot\gamma = \sqrt{(1-z)}/\overline s$. In each case, complete segregation is observed, where every large particle lies above every small particle. The non-uniform shear strain rate in the bottom case causes non-uniform transient behaviour towards a steady grading that is the same as the top case. According to the formulation presented here, the final grading is independent of the loading condition.

As shown in \cite{sizeism}, this model represents a very simple analogy of the analytic works done by \cite{savage1988pss,gray2005tps} to model bidisperse segregation, with a very simple extension to polydisperse systems, as shown below.

\begin{figure}
    \centering
    \includegraphics[width=\columnwidth]{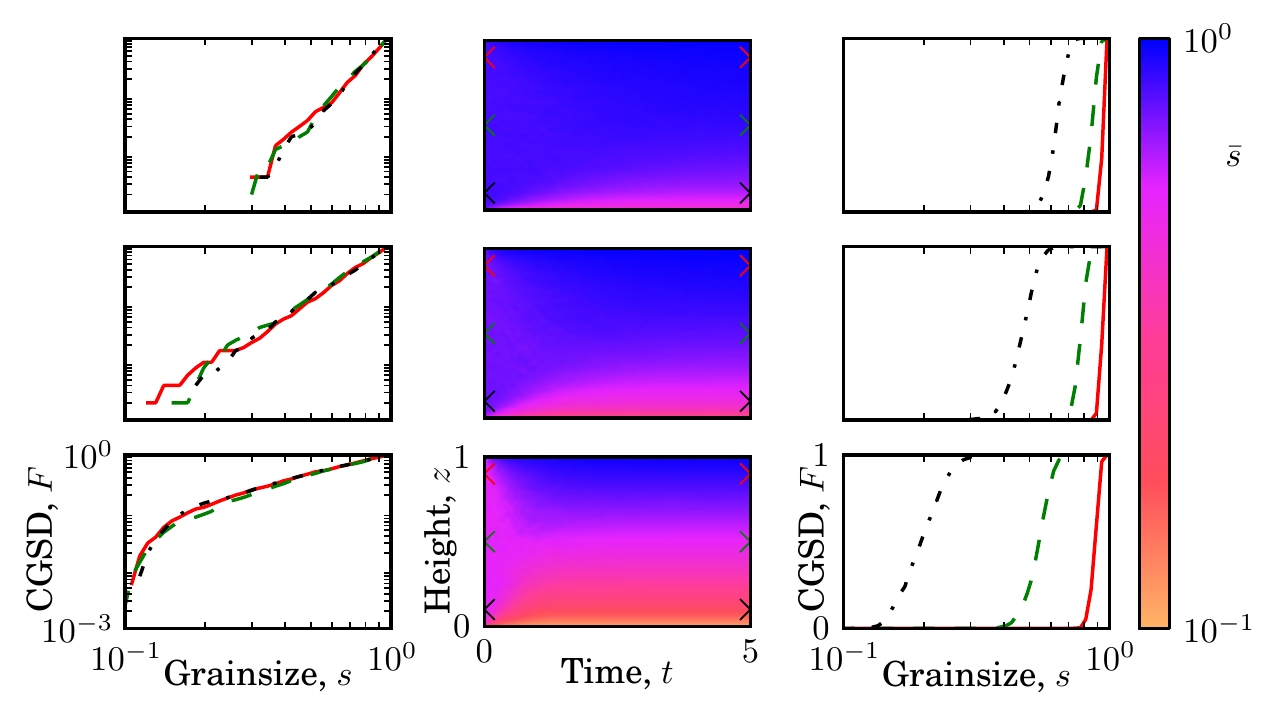}
    \caption{Polydisperse segregation under simple shear. For all cases $k_s=1$ and $\dot\gamma=1$. \emph{Top to Bottom}: Each row represents a single simulation with initial condition defined by $\alpha_i = -2$, 0 and 2 respectively. \emph{Left}: Initial cumulative grainsize distribution at three different heights. The solid red, dashed green and dash-dotted black lines represents $z=0.9$, 0.5 and 0.1 respectively. \emph{Middle}: Plot of the average grainsize $\bar s$ over height and time. \emph{Right}: Final cumulative grainsize distributions, plotted in the same manner as the initial grainsize distributions. \label{fig:poly_seg_simple}}
\end{figure}

\subsection{Polydisperse segregation}\label{sec:poly}

We can create a polydisperse sample by generating initial conditions in the same way as previously described for the breakage cellular automaton, using Equation \ref{eq:initial}. The segregation patterns produced for a range of initial conditions at constant segregation rate $k_s=1$ and with $\dot\gamma=1$ are shown in Figure \ref{fig:poly_seg_simple}. Since this is now a polydisperse sample, we can calculate the grainsize distribution $\phi(s)$. On the left hand side of Figure \ref{fig:poly_seg_simple} are the initial cumulative grainsize distributions, which are homogeneous. During the simulation, segregation occurs, creating a non-homogeneous steady state condition after some time. These grainsize distributions, which now vary with height, are shown on the right hand side of the same Figure.

Averaging over all cells at a given height $i$, we can express the mean segregative velocity $u_{i}$ of a single grainsize fraction centred at $s_a$ from all cells at height $i$ in a time $\Delta t$ as

\begin{figure}
    \centering
    \includegraphics[width=\columnwidth]{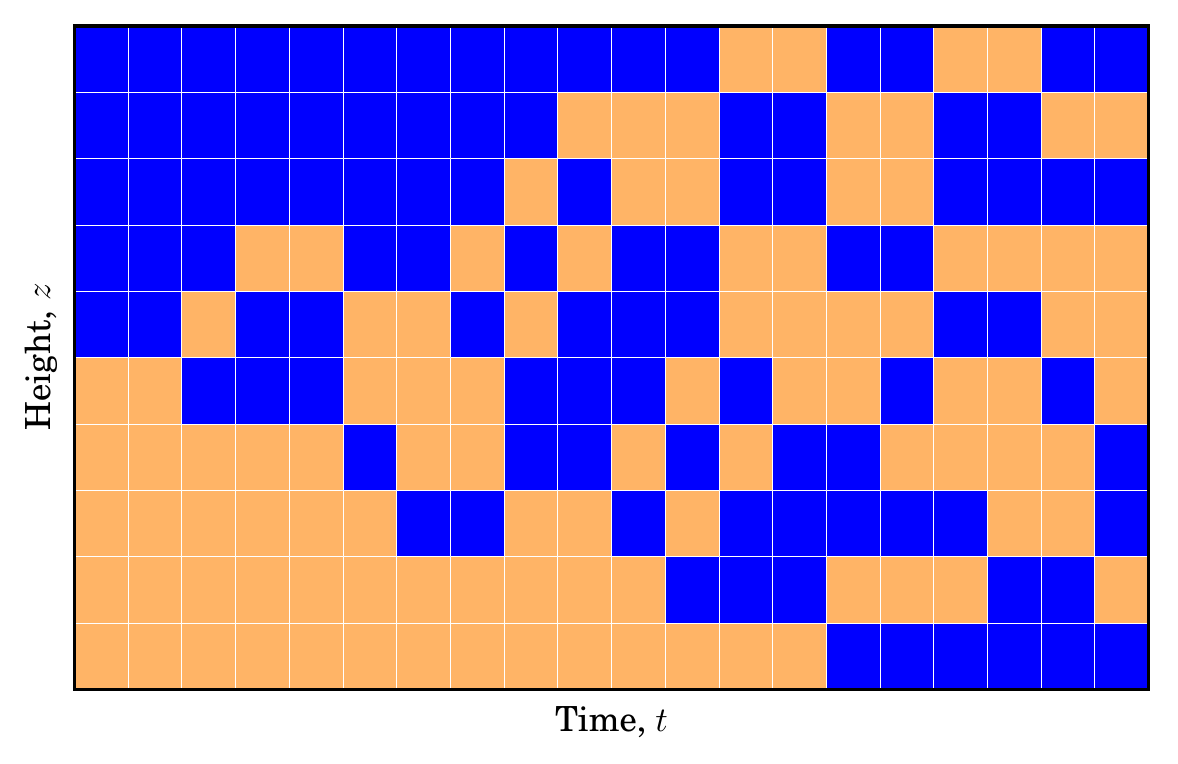}
    \caption{The mixing mechanism. Ten cells, initially segregated with all large particles (blue) above small particles (yellow), subjected to the mixing mechanism only. Over time, the system reaches a disordered state. \label{fig:mix_small}}
\end{figure}

\begin{equation}
u_{i}(s_a) = \frac{1}{N_x}\sum_{j=1}^{N_x} f_{i,j}(s_a) = \frac{k_s|\dot\gamma_{i}|}{N_x}\sum_{j=1}^{N_x} \left(\frac{s_a}{\overline s_{i,j}} - 1\right).
\end{equation}

We have included the local average grainsize $\bar s_{i,j}$ in the formulation so that we know which particles are locally small or large. As particles are being swapped between heights --- between representative volume elements at the continuum scale --- we only require a single average grainsize per height, and can in this case freely extend the neighbourhood domain over which we find the average grainsize $\overline s$ to include every cell at height $j$, labelling it now $\overline s_i = 1/N_x\sum_j s_{i,j}$. In this case, the mean velocity can be expressed as

\begin{equation}
u_{i}(s_a) = k_s|\dot\gamma_{i}|\left(\frac{s_a}{\overline s_i} - 1\right).
\end{equation}

Compare this with the analytic description of the segregation velocity with no diffusion as predicted by \cite{grainsize} for a continuum with internal grainsize coordinate $s$,

\begin{equation}
u(s) = |\dot\gamma|\frac{g\cos\theta}{c}\left(\frac{s}{\overline s} - 1\right),
\end{equation}

\noindent where $c$ is a fitting parameter, $g$ is the acceleration due to gravity, and $\theta$ is the angle of the plane down which flow is occurring. As shown in \cite{marks2011polydisperse}, cellular automata can successfully be used as a coarse finite differencing method to model systems such as these without resorting to complicated flux limited finite difference schemes, as would otherwise be necessary.

This model, together with the rule for remixing, which will be shown next, represents the simplest description of the analytic description presented in \cite{grainsize}.

\begin{figure}
    \centering
    \includegraphics[width=\columnwidth]{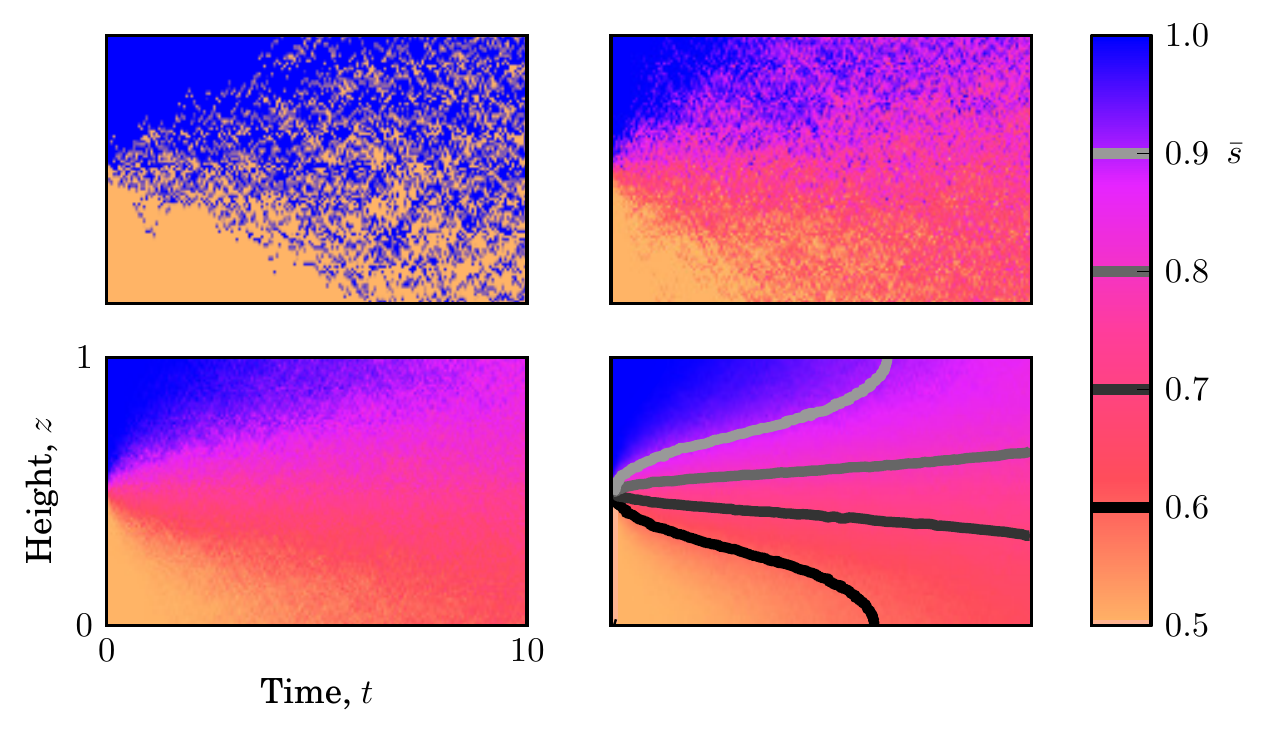}
    \caption{The time evolution the average grainsize $\bar s$ of a bidisperse simulation with varying $N_x$ subject to mixing only. All cases have $D=0.01$ and $N_z=100$. \emph{Clockwise from Top Left:}  $N_x=1$, 5, 50 and 1000. \emph{Bottom right:} Solid lines indicate contours from solution of analogous continuum equation.\label{fig:mix_averaging}}
\end{figure}

\subsection{Remixing}

In nature we rarely see such perfect segregation as that pictured above. This is due to the random fluctuation of particles as the flow propagates down slope. As has been done before analytically \cite{GrayChugunov}, we can capture this effect by introducing remixing into the flow. For the simplest case, we allow particles to swap randomly either up or down with some frequency, given by a constant $D/\Delta z^2$. At this stage we let this probability be independent of the shear strain rate $\dot\gamma$, although a strong dependency has been observed \cite{UtterBehringer} in experiments. With frequency of swapping controlled by the diffusivity, $D$,

\begin{eqnarray*}
\text{Flip a coin, if heads: } &s_i \Leftrightarrow s_{i-1}& \\
\text{if tails: } &s_i \Leftrightarrow s_{i+1}&
\end{eqnarray*}

An example of the mixing rule acting on a single column of cells over time is shown in Figure \ref{fig:mix_small}. Initially, the system is perfectly segregated, but over time the systems becomes randomised due to the presence of remixing. The characteristic time for mixing to occur is the inverse of the diffusivity $D/H^2$.

We are describing a system of cells undergoing Brownian motion, whereby particles move by the application of random forces over time scales that are short relative to the motion of the particle. When considered over long time scales and large numbers of particles, this is analogous to Fickean diffusion \cite{UtterBehringer}. Many other cellular automata exist to model pure diffusion \cite{chopard1991cellular}.

As in the case of segregation, this process can be averaged over the $x$ direction to describe the evolution of the average grainsize at any height over time. By increasing the number of cells in the $x$ direction, we can increase the smoothness of our solution. Figure \ref{fig:mix_averaging} shows the same system as Figure \ref{fig:mix_small}, initially segregated, that mixes over time to create a homogeneous system, but now for increasing numbers of cells in the $x$ direction.

This diffusive behaviour can be described at the continuous limit using Fick's first law of diffusion,

\begin{equation}\label{eq:fick}
\frac{\partial\phi}{\partial t} = D\frac{\partial^2\phi}{\partial t^2}.
\end{equation}

\begin{figure}
    \centering
    \includegraphics[width=\columnwidth]{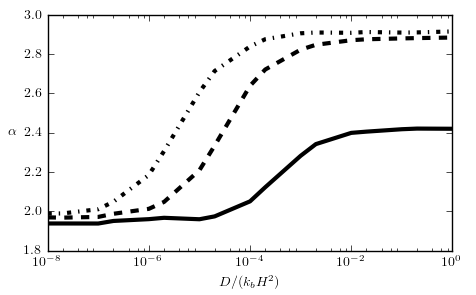}
    \caption{Coupled comminution and mixing. Varying values of $D/(k_bH^2)$. For all cases, $N_z=21$, $N_x=1000$. Plot shows final value of best fit to power law part of cumulative grainsize distribution at $t=5$, 50 and 500, corresponding to the solid, dashed and dash-dotted lines respectively. \label{fig:crush_and_mix}}
\end{figure}

\section{Coupled problems}

We now have three distinct processes which can be described simultaneously in a single simulation. These have all been shown above with their analogous continuum description, yet not in all cases could a direct link be shown. For the case of comminution, an internal length scale governing the spatial distribution of grainsize over a sub-continuum length scale was required.

As all of the mechanisms previously described have been created in the same framework, we can simply run a cellular automaton which includes multiple phenomena at the same time. As will be shown in this Section, we can investigate the interactions between the mechanisms by varying the parameters which control their effects.

\subsection{Comminution and mixing}

We begin with a cellular automata that includes both comminution and mixing. This system represents an extension of the cycles of crushing pictured in Figure \ref{fig:cycles}, but now with true advection. In this case, as in Figure \ref{fig:cycles}, we expect that after a short time relative to the diffusive time, the system will reach $\alpha_f=2$, as significant mixing has not yet occurred. At longer times, the system will approach $\alpha_f=3$, and its final grading. This effect is captured in Figure \ref{fig:crush_and_mix}, where the diffusive time is controlled by the ratio $D/(k_bH^2)$, and the system is constrained at $\alpha_f = 2.91$.

For vanishingly small diffusivities, the system will still approach $\alpha_f=3$, but only after very long periods of comminution. Conversely, at very large diffusivities, the system passes $\alpha_f=2$ very rapidly, and approaches $\alpha_f=3$ in a relatively short time.

\subsection{Segregation and mixing}

In flows of polydisperse granular materials where comminution does not occur, we can model the evolution of the grainsize distribution as being comprised of segregative and diffusive remixing components. This occurs in many industrial mixing processes, and may be sufficient to model levee formation and runout characteristics in landslides. At higher speeds remixing increases, suppressing segregation, while at low speeds segregation can play a dominant role in the flow behaviour.

\begin{figure}
    \centering
    \includegraphics[width=\columnwidth]{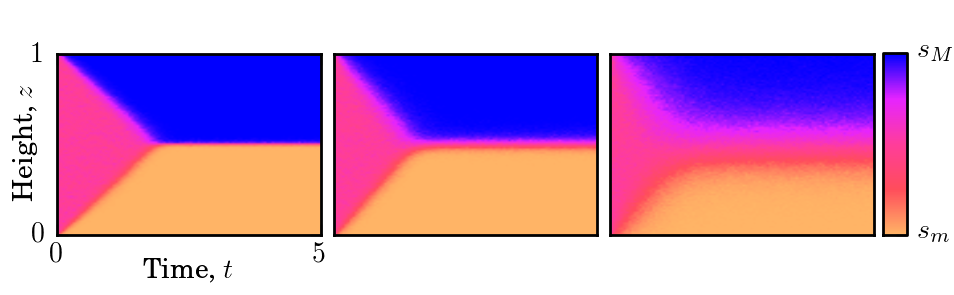}
    \caption{Bidisperse segregation under simple shear with diffusion. \emph{Left to Right}: Increasing diffusivity $D=0.002$, 0.01 and 0.05 with constant segregation coefficient $k_s=1$. Increasing the mixing coefficient smooths out concentration shocks in the spatial direction, giving more physically representative solutions.\label{fig:bi_seg_diffusion}}
\end{figure}

The effect of coupling mixing and segregation in a bimixture can be seen in Figure \ref{fig:bi_seg_diffusion}, where increasing diffusivity $D$ smooths out the concentration shock between the two phases of large and small particles. This can be treated in an identical manner for polydisperse mixtures.

This has been shown analytically for bidisperse systems in \cite{GrayChugunov} and validated experimentally in \cite{Wiederseiner2011}.

Generally, suppressing spurious numerical diffusion is a non-trivial task, requiring sophisticated finite differencing schemes \cite{Kurganov2000241} to maintain hyperbolicity. This type of stability, which generally controls the accuracy of the numerical results, is not an issue for solutions obtained using cellular automata \cite{sizeism}.

\begin{figure*}
    \centering
    \includegraphics[width=\textwidth]{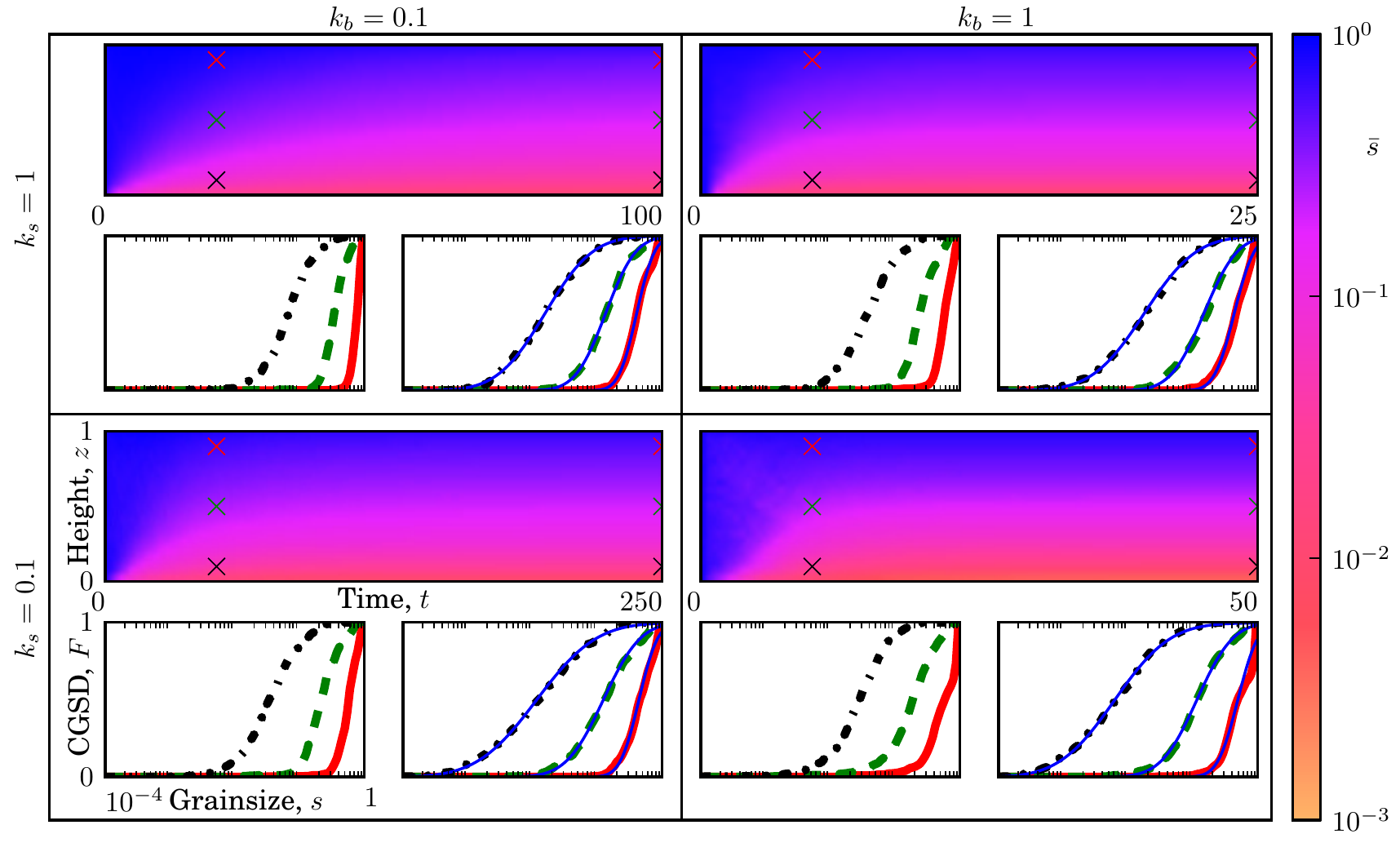}
    \caption{A crushable flow with two mechanisms: segregation and comminution. Initially, the system is homogeneous, being a polydisperse sample with initial grainsize distribution defined by $\alpha_i=-2$. Four cases are considered with varying $k_b$ and $k_s$. For each case, three plots are shown. \emph{Top}: Evolution of the average grainsize at every height over time. \emph{Bottom left}: Cumulative grainsize distribution at three points in the flow, corresponding to the crosses in the above plot. Top is red solid line, middle is green dashed line and bottom is black dash-dotted line.  \emph{Bottom right}: Cumulative grainsize distributions at steady state at same heights as previously. Solid blue lines represent best fit cumulative log-normal distributions.\label{fig:crush_and_seg}}
\end{figure*}

\subsection{Segregation and comminution}

In many situations, segregation and comminution occur simultaneously in a flow situation, such as in the grain milling depicted in Figure \ref{fig:mill_stone}. In other cases, it is not even clear if segregation has occurred, yet particles are advecting in space and strong comminution is observed, such as in earthquake faulting and snow avalanches. In many of these cases, we observe log-normal grainsize distributions, rather than power law distributions, which exist at all depths of flow. The existence of these curves represents the competition between the two mechanisms, where the comminution attempts to form a power law distribution, and the segregation attempts to create a locally monodisperse distribtion. A log-normal distribution is one that obeys the following scaling for the cumulative grainsize distribution $F_{LN}$,

\begin{equation}
F_{LN} = \frac{1}{2}\text{erfc}\left(-\frac{\ln s - \mu}{\sigma\sqrt{2}}\right),
\end{equation}

\noindent where $\mu$ and $\sigma$ are the location and scale parameters, and erfc is the complimentary error function.

In Figure \ref{fig:crush_and_seg}, simulations are shown in which both segregation and comminution are present. For all cases, log-normal cumulative grainsize distributions are observed over all depths, with p-values in the bottom half generally less than 0.001.

As expected, increasing $k_s$ or $k_b$ decreases the time to reach a steady state in terms of the average grainsize $\bar s$. It is evident that significant changes in the grainsize distribution will not occur indefinitely, even though segregation brings together particles of similar size, and comminution is accelerated by the segregation.

\subsection{Segregation, mixing and crushing}

We can now couple all three mechanisms and observe the evolution of the grainsize distribution as all of the constituent mechanisms interact. Each time step, we first check each cell and if the breakage rule is met, we change the cell's grainsize. Secondly, we iterate over all of the cells and swap them with a neighbour if the segregation rule is met. Finally, we again iterate over all cells and if the diffusion rule is met, we swap randomly with a neighbour.

\begin{figure}
    \centering
    \includegraphics[width=\columnwidth]{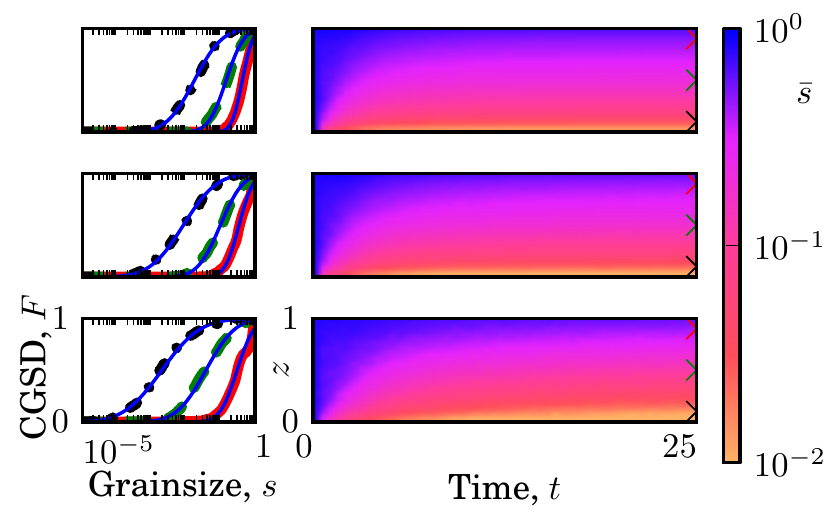}
    \caption{A crushable flow with all three mechanisms. For all cases $k_s=k_b=1$ and $H=1$ is the height of the system. Initially, the system is homogeneous, being a polydisperse sample with initial grainsize distribution defined by $\alpha_i=-2$. Each row represents the same initial condition but with varying amounts of mixing. From top to bottom, $D=0$, 0.005 and 0.05 respectively. \emph{Left}: Cumulative grainsize distribution averaged at three heights $z=0.1$, 0.5 and 0.9 at steady state (black, green and red respectively). \emph{Right}: The average grainsize at every height evolving over time. \label{fig:sturzstrom}}
\end{figure}

We now have a system that models avalanche and landslide flow, where particles at the base are sheared and crush, creating a lubrication layer, such as in Figure \ref{fig:crush_and_seg}. The inclusion of mixing in the system, as shown in Figure \ref{fig:sturzstrom}, enhances the spread of sizes produced by comminution, and reduces the size of particles in the bottom-most layer of flow, enhancing the lubrication effect. Again, log-normal cumulative grainsize distributions are measured, which represent those found in many geophysical processes, such as in snow avalanches \cite{BarteltGranulometric2009}.

%Since we observe log-normal cumulative grainsize distributions in most simulations with coupled segregation, mixing and comminution, it is possible to show how the parameters of these distributions, $\mu$ and $\sigma$, and their depth dependency, can be related to the coefficients $k_s$, $k_b$ and $D$.

\subsection{An equivalent continuum model}

Considering conservation of mass alone, we can express all three mechanisms in a continuum form as

\begin{eqnarray}
\frac{\partial\phi}{\partial t} + k_s\frac{\partial}{\partial z}\left(\phi|\dot\gamma|\left(\frac{s}{\bar s} - 1\right)\right) = D\frac{\partial^2\phi}{\partial t} \nonumber\\
+ b\phi - \int_s^{s_M}P(s|s\rq{})b(s\rq{})\phi(s\rq{})~ds\rq{}.
\end{eqnarray}

This model differs from the cellular automata in the sense that the internal coordinate $s$ does not retain information about local neighbours. Because of this, cycles of crushing cannot be produced. This omission gives an important insight into the use of continuum theories to represent internally (i.e.\ within the representative volume element) spatially correlated material.

For the mechanisms of segregation and mixing, the length scale representing the local neighbourhood is not an important consideration. However for comminution, it must be included as part of the model to enable us to predict the correct final distribution.

\section{Conclusions}

We have shown that cellular automata can successfully capture the most important physical foundations behind evolving phenomena in crushable granular flows. To do this, we have used three distinct cellular automata to explain the dominant mechanisms during such flows: comminution, segregation and mixing.

By assembling all three cellular automata together we were able to explore the interactions between these phenomena. One surprising outcome is that in closed systems, crushable granular material are limited by power laws, however during flow the interaction with segregation and remixing the system is limited by log-normal distributions.

This paper highlights the power of the cellular automata as a means to inspire continuum models. We have demonstrated that it is often possible to recover an analogous continuum description from the limit of the cellular automata rules. An interesting result of this tactic is the result that comminution does not follow this rule. For comminution, the ability to model cycles of crushing using a conventional continuum model with a grainsize coordinate is currently impossible, as some local rules are inherent to the system that must be included to describe the local configuration of grains.

The success of this cellular automata is that it enables us to study the evolution and limits of the grainsize distribution in different scenarios. Current continuum models require a priori knowledge the grainsize distribution, and cannot educate us on the physical mechanisms involved in reaching this final state.

IE acknowledges grant DP0986876 from the ARC.

\bibliographystyle{spphys}       % APS-like style for physics

\end{document}